\documentclass[prl,aps,twocolumn,tightlines,showpacs]{revtex4}

\usepackage[]{times,amsmath,epsfig,graphicx}

\begin{document}

\title{Optical Kerr Frequency Comb Generation in Overmoded Resonators}

\author{ A. B. Matsko, A. A. Savchenkov, W. Liang, V. S. Ilchenko, D. Seidel, and L. Maleki}

\affiliation{OEwaves Inc., 465 N. Halstead Street, Pasadena, CA 91107 }

\begin{abstract}
We show that scattering-based interaction among nearly degenerate optical modes is the key factor in low threshold generation of Kerr frequency combs in nonlinear optical resonators possessing small group velocity dispersion (GVD). The mode interaction is capable of producing drastic change in the local GVD, resulting in either a significant reduction or increase of the oscillation threshold. It is also responsible for the majority of observed combs in resonators characterized with large normal GVD. We present results of our numerical simulations as well as supporting experimental data.
\end{abstract}

\pacs{42.62.Eh, 42.65.Hw, 42.65.Ky, 42.65.Sf}

 \maketitle

There are several experimentally observed phenomena in optical frequency combs generated with monolithic microresonators (Kerr combs) that have been studied by multiple researchers (see \cite{kippenberg11s} for review),  but are not as yet explained theoretically. An example of the observed phenomena lacking a theoretical explanation is the  strong dependence on the selection of the externally pumped optical mode of both the spectral envelope and the generation threshold of the comb. The purpose of this contribution is to provide a theoretical framework for these observations.

Generation of optical frequency (Kerr) combs results from modulation instability of the continuous wave (cw) light confined within a mode of the resonator \cite{matsko05pra}. The comb is generated when the power of the cw pump exceeds a certain threshold. The frequency harmonics of the comb are produced in the modes of the resonator adjacent to the pumped mode; these harmonics are phase locked \cite{delhaye07n}.

Theoretical studies dealing with idealized nonlinear resonators suggest that the comb spectra are symmetric with respect to the frequency of the optical pump \cite{agha09oe,agha07pra,chembo10prl,chembo10pra}. Furthermore, the frequency spacing of the generated harmonics can vary from a single free spectral range (FSR) (the fundamental comb) to multiple FSRs of the resonator, depending on the power and frequency of the pump as well as group velocity dispersion (GVD) of the resonator modes \cite{savchenkov08prl,arcizet09chap}.

Theoretical predictions  \cite{agha09oe,agha07pra,chembo10prl,chembo10pra} describe results of experiments performed with resonators having anomalous GVD \cite{liang11ol}, while multiple experimental observation with normal or nearly-zero GVD resonators \cite{savchenkov08prl,grudinin09ol} evidently contradict the theory. In resonators characterized with anomalous GVD the power-dependent frequency shift of the modes is compensated by the dispersion, leading to extended dynamic range for comb formation as well as ensuring excitation of the comb from zero field fluctuations (soft excitation regime). Theoretically, there is no soft excitation regime for combs of resonators with normal GVD and the dynamic range for the comb formation is limited there \cite{matsko11nlo}. Practically, asymmetric combs and the soft excitation regime of the comb are frequently observed. In this work we show that the discrepancy originates from the interaction of resonator modes overlapping the same mode volume and belonging to different mode families. The mode interaction disturbs the regular GVD of the resonator spectrum so much that the local GVD significantly changes its value, as well as its sign. We argue that  mode crossing is the chief reason for observation of low threshold frequency combs in large enough monolithic resonators.

Very recently performed a study of universal dynamics of Kerr comb formation in dielectric microresonators \cite{herr11arch}. This study explained the anomalously large linewidth broadening in octave spanning frequency combs \cite{delhaye11prl} and multiple beat-notes observed in low repetition rate comb systems when the comb was demodulated on a photodetetor \cite{papp11pra}. Our current study shows that the theoretical consideration of the low repetition rate comb systems should be adjusted to take into account the interaction between the modes of the nonlinear resonator.

To reveal the importance of mode crossing we numerically simulate generation of combs in 21 identical optical modes coupled through cubic nonlinearity. The modes have unequal separation resulting from nonzero GVD of the resonator. To perform the simulation we numerically solve a set of nonlinear equations derived using the physical model presented in \cite{matsko05pra,chembo10pra}:
\begin{equation} \label{set}
\dot{\hat a}_j=-(\gamma_0+i\omega_j)\hat a_j+ \frac{i}{\hbar} [\hat V,\hat a_j]+F_0 e^{-i\omega t} \delta_{11,j},
\end{equation}
where $\hat V= -\hbar g (\hat e^\dag)^2 \hat e^2/2$ is the interaction Hamiltonian, $\hat e= \sum _{j=1}^{21} \hat a_j$ ($\hat a_j$ is the annihilation operator of the field in the mode), $g=\hbar \omega_{11}^2 c n_2/({\cal V}n_0^2)$ is the interaction constant, $\omega_{11}$ is the frequency of the pumped mode, $n_0$ and $n_2$ are the linear and nonlinear refractive indexes of the material, ${\cal V}$ is the mode volume, $\delta_{11,j}$ is the Kronecker's delta; $\gamma_0=\gamma_{0c}+\gamma_{0i}$ is the half width at the half maximum for the optical modes, assumed to be the same for the all modes involved; and $\gamma_{0c}$ and $\gamma_{0i}$ stand for coupling and intrinsic loss. The external optical pumping is given by $F_0= (2 \gamma_{0c} P/(\hbar \omega_{11}))^{1/2}$, where $P$ is the value of the cw pump light. We neglect the quantum effects and do not take into account corresponding Langevin noise terms.

We solved (\ref{set}) numerically taking into account only the second order frequency dispersion recalculated for the frequency of the modes $\omega_j$ as $\omega_{12}+\omega_{10}-2 \omega_{11} = -\beta_2 c\omega_{FSR}^2/n_0$, where $\beta_2$ is the effective GVD of the modes with no interaction taken into account, $c$ is the speed of light in the vacuum, $\omega_{FSR}$ is the free spectral range of the resonator, e.g. $2\omega_{FSR} \simeq \omega_{12}-\omega_{10}$.

The effective GVD value is negligible in a large WGM resonator that has no mode interaction, $\gamma_0 \gg |D|=| \omega_{12}+\omega_{10}-2 \omega_{11}|$. For example, the mode non-equidistance can be estimated as $D \simeq -2\pi \times 39$~Hz ($D \simeq 2 \pi \times 1$~kHz) in a 10~GHz FSR CaF$_2$ (MgF$_2$) WGM resonator pumpeded at 1550~nm. The value changes rather insignificantly compared with the typical bandwidth of the modes used in reported experiments ($2 \gamma_0 > 2 \pi \times 200$~kHz, or $Q<10^9$) in smaller resonators (35~GHz FSR): $D \simeq -2\pi \times 5.2$~kHz ($D \simeq 2 \pi \times 8$~kHz). On the other hand, $D \approx 2 \gamma_0$ is required for generation of the fundamental frequency comb with soft excitation \cite{matsko05pra}. This is possible in a small resonator. For instance, for a fundamental TE mode of a 100~GHz FSR MgF$_2$ WGM  resonator pumped with 1721~nm light $D \approx 2 \pi \times 200$~kHz. Any observation of soft excitation of the comb in a larger resonator is the evidance that the spacing of the modes is somehow altered compared with the value of an ideal resonator. One way for such a modification is related to the usage of a different mode family, the GVD of which can be tuned with the morphology of the resonator \cite{savchenkov11np}. Another way is related to the mode interaction described below.

Interaction between resonator modes has been encountered in WGM resonators. It is directly observed as a disruption of the continuity of dispersion in smaller resonators \cite{delhaye09np}, and is indirectly indicated by the asymmetry of the Kerr comb spectra \cite{savchenkov08prl,grudinin09ol} (the spectrally narrow combs generated in resonators with no mode interaction has to be symmetric because of energy conservation). A spectrum of a microresonator taken in a broad frequency range \cite{ferdous11np} clearly shows the presence of  spurious modes changing their position with wavelength with respect to the fundamental mode family. Since the spurious and the fundamental modes are able to interact due to unavoidable imperfections of the resonator shape, they also can change the dispersive properties of the resonator. Even if the spurious modes are not seen in the experiment with a particular selection of the coupling technique, they still can exist in the resonator \cite{savchenkov07pra,carmon08prl}. Only a resonator supporting a single mode family (see, e.g., \cite{savchenkov06ol}) can be considered free from mode interaction. Therefore, it is natural to expect that mode interaction is essential in the majority of observations of Kerr frequency combs in optical resonators.

\begin{figure}[htbp]
  \centering
  \includegraphics[width=8.5cm]{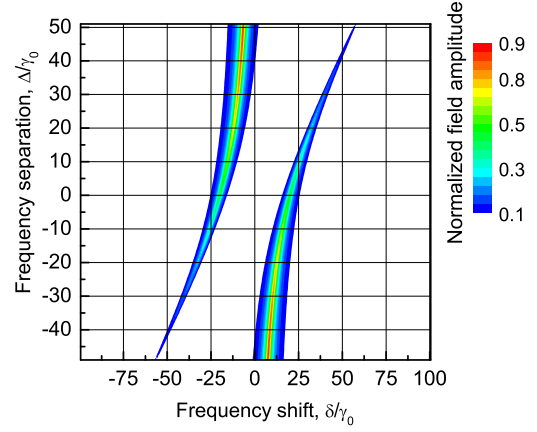}
\caption{Density plot illustrating frequency and amplitude dependence of the forced response of a linear mode $a_{12}$ on the detuning $\Delta= \omega_c-\omega_{12}$. Mode $a_{12}$ is pumped externally with a monochromatic force of a constant amplitude ($|F_0/\gamma_0|=1$). The frequency separation between the force frequency ($\omega$) and the frequency of the mode ($\tilde \omega_{12}$) is determined as $\delta=\tilde \omega_{12}-\omega_{12}$, where $\omega_{12}$ corresponds to the frequency of the free mode. The calculations are made for $\kappa/\gamma_0=20$. The dependence shows that for large $|\Delta|$ the interaction primarily results in frequency shift of the mode of interest. The mode splits into a symmetric doublet when $\Delta$ approaches zero.} \label{fig1}
\end{figure}

Let us assume that the mode containing the first higher frequency harmonic ($a_{12}$) interacts with a mode of the resonator ($c$) having the same loaded Q-factor and not belonging to the mode family the comb is generated in. The interaction, described by Hamiltonian $\hbar \kappa (a_{12}^\dag c+c^\dag a_{12})$, results in the well known splitting of the resonance for $a_{12}$ (see Fig.~\ref{fig1}). The splitting may be considered as a pure frequency shift of the mode $a_{12}$ expressed as
\begin{equation} \label{shift}
\tilde \omega_{12}=\omega_{12}-\frac{\kappa^2}{\Delta},
\end{equation}
in the asymptotic case of large difference between the eigenfrequencies of the interacting modes, $\Delta = \omega_c-\omega_{12}$, compared with the interaction constant $\kappa$, $|\Delta| \gg \kappa$.

The shift of mode $a_{12}$ resulting from the interaction with mode $c$ leads to the modification of the effective GVD value for the pumped mode and the first two sideband modes. Really, using Eq.~(\ref{shift}) we find
\begin{equation} \label{mod_dispersion}
\tilde \omega_{12}+\omega_{10}-2 \omega_{11} = -\beta_2 \frac{c\omega_{FSR}^2}{n_0}-\frac{\kappa^2}{\Delta}.
\end{equation}
According to  Eq.~(\ref{mod_dispersion}), it is enough to have $\kappa^2 \approx -2 \Delta \gamma_0$ to achieve the desirable value of the GVD in any resonator with a small intrinsic dispersion.
\begin{figure}[htbp]
  \centering
  \includegraphics[width=8.5cm]{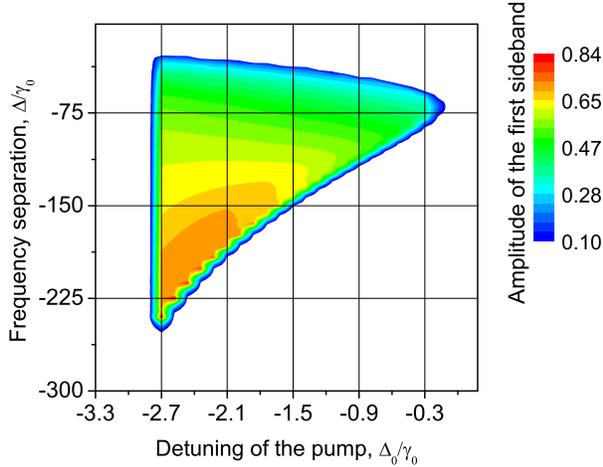}
\caption{ Density plot showing normalized amplitude distribution of the first higher frequency sideband of the Kerr comb generated due to interaction of the resonator modes. The distribution is plotted as a function of the detuning between the pumping laser and the corresponding mode of the resonator ($\Delta_0/\gamma_0$), and the frequency separation between the optical mode containing the comb harmonic and the service mode ($c$). Coupling parameter $\kappa/\gamma_0=20$ is selected. The GVD of the resonator is selected to be $D/\gamma_0=-0.02$. The amplitude of the external cw pump is $|F_0|/\gamma_0=2$. Mode $a_{11}$ is pumped.
} \label{fig2}
\end{figure}

\begin{figure}[htbp]
  \centering
  \includegraphics[width=8.5cm]{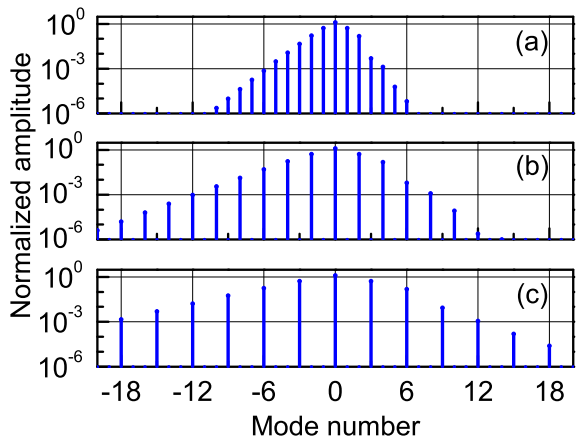}
\caption{ Spectra of the frequency combs generated in a resonator having nearly zero normal GVD ($D/\gamma_0=-0.02$) when one of the resonator modes is shifted due to the interaction with another mode. Coupling parameter $\kappa/\gamma_0=20$ is selected. The amplitude of the external cw pump of the central mode ($a_{11}$) is $|F_0|/\gamma_0=2$. Three cases are shown: (a) mode $a_{12}$ is shifted due to the interaction; (b) mode $a_{13}$ is shifted; and (c) mode $a_{14}$ is shifted. Because the relative GVD of the resonator is small it does not influence the nonlinear process, so shapes of the comb envelopes simply scale from picture to picture.
} \label{fig3}
\end{figure}

To validate the analytical calculations we performed a simulation for the 21 mode at the above conditions but also took into account the interaction with a mode $c$. Selecting $\kappa/\gamma_0=20$ we found that the soft excitation of the Kerr comb is possible for a wide range of frequency detunings: $-0.2 \gamma_0>\Delta>-2.7\gamma_0$ (Fig.~\ref{fig2}). The Kerr comb for specifically selected parameters of the system is shown in Fig.~(\ref{fig3}a). It has a slightly asymmetric spectrum and fast roll-off of the higher order harmonics. The comb starts from zero fluctuations of the field with essentially zero initial conditions. Since such excitation regime of the comb is absent in the case where no mode interaction is available, we conclude that mode interaction is the cause, as predicted by the reasoning above.

\begin{figure}[htbp]
  \centering
  \includegraphics[width=8.5cm]{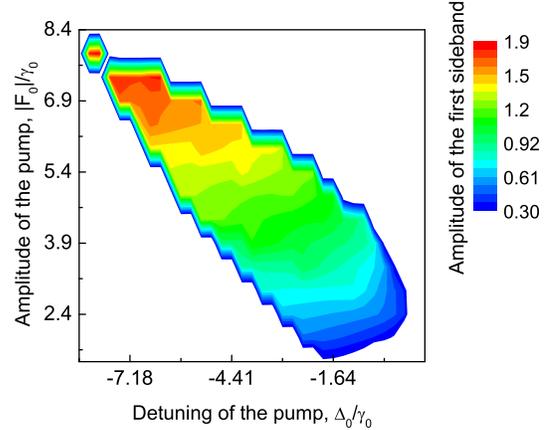}
\caption{ Normalized amplitude of the first comb sideband generated in the mode $a_{12}$ shifted due to the interaction with the service mode. Nearly zero normal GVD ($D/\gamma_0=-0.02$) and coupling parameter $\kappa/\gamma_0=20$ are selected.
} \label{fig4}
\end{figure}

\begin{figure}[htbp]
  \centering
  \includegraphics[width=8.5cm]{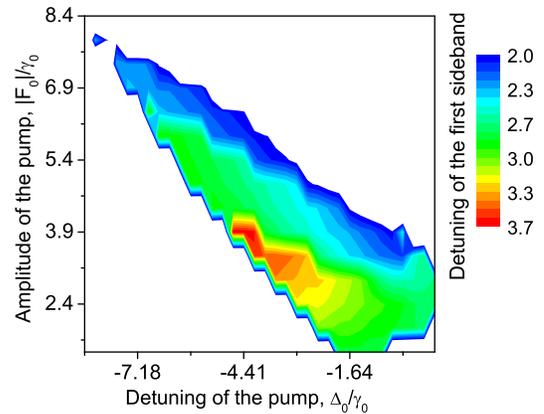}
\caption{ Frequency difference between the local FSR of the resonator (not modified by the interaction with the service mode) and the repetition rate of the comb as a function of the amplitude and frequency of the pump.
} \label{fig5}
\end{figure}

According to the numerical simulations the frequency comb based on mode interaction has two distinct features: (i) the dynamic range of the soft excitation of the comb is limited with respect to the power of the external pump and (ii) the repetition rate of the comb depends on the pump power. We evaluated the power of the first comb sideband generated in the shifted resonator mode ($a_{12}$) as a function of the pump amplitude and frequency detuning for the fixed interaction value with another mode (Fig.\ref{fig4}). The limitation occurs due to clamping of the value of the nonlinear frequency shift of the resonator modes that can be compensated by the interaction of $a_{12}$ and $c$ modes. The power dependence of the comb frequency is illustrated by (Fig.\ref{fig5}), showing the difference $[(\omega_{12}-\omega_{11})-({\omega'}_{12}-\omega)]/\gamma_0$, where ${\omega'}_{12}$ is the frequency of the generated sideband.

Interaction of the modes also explains experimentally observed generation of combs having different repetition rates in resonators with normal GVD \cite{savchenkov08prl}. The comb step changes depending which of the comb-generating mode in the sequence is shifted due to the interaction. The intrinsic GVD of the resonators is too small to influence the process. We have simulated generation of higher repetition rate combs and validated the concept (Fig.~\ref{fig3}b,c). The simulation shows that the detuning of the pump light from the corresponding mode of the resonator, discussed in \cite{savchenkov08prl}, is of less importance  for generation of high repetition rate frequency combs having soft excitation regime, compared with mode interaction considered here. The change of the comb step in normal GVD resonators is different from the mechanism of the step change in resonators with anomalous GVD, where repetition rate strongly depends on the intracavity power of the pump light \cite{arcizet09chap,liang11ol}.

\begin{figure}[htbp]
  \centering
  \includegraphics[width=8.5cm]{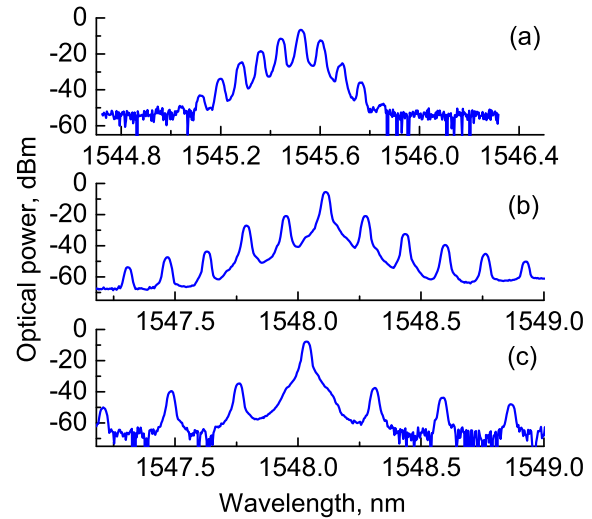}
\caption{ Experimental spectra of three optical frequency combs generated in the same overmoded WGM resonator when light is coupled to three arbitrary modes having strong interaction with other modes. The wavelength of the pumping light was changed to select the modes, and depending on the mode selection we observed (a) single-FSR, (b) dual-FSR comb, and (c) triple-FSR combs. Generation of higher order frequency combs, not shown here, was also recorded.
} \label{fig6}
\end{figure}

We performed an experiment with a large WGM resonator having nearly zero relative GVD ($D/\gamma_0$) and observed generation of  frequency combs with envelopes similar to those predicted by the theory. In our experiment we used a CaF$_2$ resonator having $6721\ \mu$m in diameter. The resonator had approximately 9.9~GHz FSR with loaded quality factor exceeding $10^9$. We pumped the resonator with 1545.5~nm light emitted by a distributed feedback semiconductor laser. The light was coupled to the resonator via a coupling prism. The optical power emitted by the laser was 15~mW, and 3.2-1~mW of the light entered the selected modes of the resonator (the value depends on the selected mode). The output light was collected using a PM Panda fiber and introduced to an optical spectrum analyzer. The resultant optical spectra are shown in Fig.~(\ref{fig6}).

Frequency combs simulated numerically have frequency shapes similar to the combs observed experimentally (compare Fig.~\ref{fig3} and \ref{fig6}). The comb properties change significantly when we pump different modes of the resonators. Such a modification cannot be explained by the change in the geometrical part of the GVD of the modes. The interaction with the degenerate modes, on the other hand, perfectly explains the observation.

To conclude, we have shown theoretically that experimentally observed generation of optical Kerr frequency combs in nonlinear resonators with small group velocity dispersion results from the linear interaction of resonator modes. The mode interaction changes the frequency of the modes enabling soft excitation regime of the frequency combs. Only  combs produced via hard excitation can otherwise be generated in these resonators.

The authors acknowledge partial support of the reported study by DARPA's IMPACT program.

\end{document}